\documentclass[aps,prd,preprint,superscriptaddress,amsmath,amssymb,showpacs]{revtex4-1}
\usepackage{dcolumn}
\usepackage{graphicx}
\usepackage{float}
\usepackage{physics}
\usepackage[colorlinks=true,allcolors=blue]{hyperref}

\begin{document}

\title{Holographic light-quark energy loss in a spinning plasma}
\author{Yan-Qing Zhao}
\email{zhaoyanqing@hainnu.edu.cn}
\affiliation{College of Physics and Electronic Engineering, Hainan Normal University,\\ Haikou 571158, China}
\author{Zhou-Run Zhu}
\email{zhuzhourun@zknu.edu.cn}
\affiliation{School of Physics and Telecommunications Engineering, Zhoukou Normal University,\\ Zhoukou 466001, China}

\begin{abstract}
In this work, we investigate light-quark energy loss in a strongly coupled plasma described by a spinning black-brane background obtained from the large-black-hole limit of the Myers-Perry geometry. The parameter $a$ characterizes the boost/rotation parameter of the dual fluid in this holographic
setup and is related to the angular velocity in the corresponding limit. We employ two complementary probes, the falling-string and shooting-string descriptions, to compute the stopping distance and the instantaneous energy loss of a light quark moving either transverse or parallel to the local tangential-flow direction. The results show that light quarks quench more readily in a hot environment. For parallel propagation, a counter-propagating configuration enhances energy loss, whereas a co-propagating configuration suppresses it. For transverse propagation, increasing the magnitude of the rotational motion promotes energy loss. In addition, the instantaneous energy loss increases along the trajectory of the endpoint. These results describe classical probe dynamics in the local planar limit and do not include intrinsic-spin effects.

\end{abstract}

\maketitle

\section{Introduction}\label{sec:01_intro}

Heavy-ion collision experiments have provided strong evidence that a quark-gluon plasma (QGP) forms under extreme conditions \cite{Arsene:2004fa,Adcox:2004mh,Back:2004je,Adams:2005dq}. One of the most informative probes of the QGP is jet quenching: energetic partons produced at the early stage of the collision lose energy while propagating through the hot and dense medium, resulting in the suppression of high-transverse-momentum hadrons \cite{Wang:1992qdg,Baier:1996kr,Majumder:2010qh,Qin:2015srf}.  Experimental data further indicate that the produced QGP behaves as a strongly coupled, nearly perfect fluid with a small ratio of shear viscosity to entropy density \cite{Shuryak:2004cy,Kovtun:2004de}. This strongly coupled nature makes perturbative methods insufficient in certain regimes and motivates the use of nonperturbative approaches.

The AdS/CFT correspondence \cite{Maldacena:1997re,Witten:1998qj,Gubser:1998bc} offers a useful theoretical framework for studying strongly coupled plasma dynamics. In this framework, a finite-temperature strongly coupled plasma is described holographically by a black hole geometry in asymptotically anti-de Sitter spacetime. Holographic methods have been extensively used to investigate parton energy loss and jet quenching in strongly coupled media \cite{Casalderrey-Solana:2011dxg,DeWolfe:2013cua}. Several complementary observables and descriptions have been developed, including the jet quenching parameter \cite{Liu:2006ug,Liu:2006he}, falling-string picture \cite{Chesler:2008uy,Arnold:2010ir,Arnold:2011qi,Chesler:2008wd}, and the shooting-string picture \cite{Ficnar:2013wba,Ficnar:2013qxa}. The falling-string and shooting-string approaches are particularly useful for describing the energy loss and stopping distance of light quarks in a strongly coupled plasma \cite{Zhu:2019ujc,Zhang:2023kzf,Zhang:2019gki,Zhang:2019jfq,Zhang:2025wxi}.

In non-central heavy-ion collisions, the produced medium can carry a large orbital angular momentum, part of which may be transferred to the QGP, giving rise to a rotating, vortical system \cite{Liang:2004ph,Becattini:2007sr,Baznat:2013zx,STAR:2017ckg,Jiang:2016woz}. This has motivated increasing interest in the effects of rotation on deconfinement, heavy-quark dynamics, and parton energy loss \cite{Chen:2020ath,Zhao:2022uxc,Golubtsova:2021agl,Arefeva:2020jvo,Golubtsova:2022ldm}. In holography, rotating black hole backgrounds
provide a natural setting in which to model such effects \cite{Hawking:1998kw,Gibbons:2004ai,Gibbons:2004js,Garbiso:2020puw,Amano:2023bhg}. Previous studies have
examined the thermodynamics of heavy quarks in spinning black hole backgrounds \cite{Zhu:2024uwu,Zhu:2024dwx,Zhu:2025ucq}, and the jet-quenching parameter in such backgrounds has also been
investigated \cite{Zhu:2025bom}.  However, most existing studies have focused either on heavy-quark observables or on the jet-quenching parameter. A systematic analysis of light-quark energy loss using both the falling-string and shooting-string approaches in a rotating background has not yet been carried out.

In this work, we study the holographic energy loss of a light quark in the local planar limit of the spinning Myers–Perry geometry, where the global rotational motion is represented by a local tangential plasma flow. The Myers-Perry black hole has a compact boundary with topology $S^3 \times \mathbb{R}$ \cite{Garbiso:2020puw} and is dual to a gauge
theory defined on this curved background.  In the large black hole limit, the geometry locally reduces to a boosted Schwarzschild black brane, where the boost parameter is related to the rotation of the dual fluid. In this planar limit, the background is homogeneous and retains only the local tangential flow of the rotating plasma. The global rotation axis and the dependence on the distance from that axis are not retained. Consequently, spatial velocity gradients, fluid vorticity, centrifugal acceleration, and possible position-dependent temperature effects are beyond the scope of the present approximation. These effects would require the full Myers-Perry geometry or another nonuniform rotating background. Although the connection to QCD is not literal, this construction provides a holographic model for exploring how rotation affects the dynamics of a strongly coupled plasma. By working in a local patch of the boundary fluid, one can isolate the kinematic effect of the local tangential flow inherited from the global rotation and study the corresponding modification of light-quark energy loss. We analyze this problem using both the falling-string and shooting-string methods. In particular, we examine how the rotation parameter affects the light-quark stopping distance and the energy loss rate. This comparison allows us to clarify the role of rotation in light-quark energy loss from two complementary holographic perspectives.

Throughout this work, light quark is used in the conventional sense of holographic energy-loss models employing classical probes. The calculation does not resolve the intrinsic spin of the probe and does not include spin–rotation or spin–vorticity couplings. The results describe stopping and energy dissipation within this approximation and do not constitute a complete description of spin-dependent light-quark dynamics in a vortical plasma.

The paper is organized as follows. In Sec.~\ref {sec:02}, we review the Myers-Perry black hole background. In Sec.~\ref {sec:03}, we study the light-quark energy loss in the spinning background using the falling-string and shooting-string approaches. In Sec.~\ref{sec:04}, we present the conclusions and discussion.

\section{Spinning Myers--Perry black hole background}\label{sec:02}

We begin by briefly reviewing the spinning black hole background considered by Hawking et al. \cite{Hawking:1998kw}, whose metric takes the form
\begin{equation}
\label{eqc1}
\begin{split}
ds^{2} & =-\frac{\Delta}{\rho^{2}}\left(dt_{H}-\frac{a\sin^{2}\theta_{H}}{\Xi_{a}}d\phi_{H}-\frac{b\cos^{2}\theta_{H}}{\Xi_{b}}d\psi_{H}\right)^{2}
 +\frac{\Delta_{\theta_{H}}\sin^{2}\theta_{H}}{\rho^{2}}\left(adt_{H}-\frac{r_{H}^{2}+a^{2}}{\Xi_{a}}d\phi_{H}\right)^{2}\\
 &+\frac{\Delta_{\theta_{H}}\cos^{2}\theta_{H}}{\rho^{2}}\left(bdt_{H}-\frac{r_{H}^{2}+b^{2}}{\Xi_{b}}d\psi_{H}\right)^{2}+\frac{\rho^{2}}{\Delta}dr_{H}^{2}
  +\frac{\rho^{2}}{\Delta_{\theta_{H}}}d\theta_{H}^{2}\\
 &+\frac{1+\frac{r_{H}^{2}}{R^{2}}}{r_{H}^{2}\rho^{2}}\left(abdt_{H}-\frac{b\left(r_H^{2}+a^{2}\right)\sin^{2}\theta_{H}}{\Xi_{a}}d\phi_{H}-\frac{a\left(r_H^{2}+b^{2}\right)\cos^{2}\theta_{H}}{\Xi_{b}}d\psi_{H}\right)^{2},
 \end{split}
\end{equation}
with
 \begin{equation}
\label{eqc11}
\begin{split}
\Delta&=\frac{1}{r_{H}^{2}}(r_{H}^{2}+a^{2})(r_{H}^{2}+b^{2})\left(1+\frac{r_{H}^{2}}{R^{2}}\right)-2M,\quad
 \Delta_{\theta_{H}}=1-\frac{a^{2}}{R^{2}}\cos^{2}\theta_{H}-\frac{b^{2}}{R^{2}}\sin^{2}\theta_{H},\\
 \rho^2&=r_{H}^{2}+a^{2}\cos^{2}\theta_{H}+b^{2}\sin^{2}\theta_{H},\quad
 \Xi_{a}=1-\frac{a^{2}}{R^{2}},\quad
\Xi_{b}=1-\frac{b^{2}}{R^{2}},
 \end{split}
\end{equation}
where $\phi_H$, $\psi_H$, and $\theta_H$ are the Hopf angular coordinates, while $t_H$, $R$, and $r_H$ denote time, the AdS radius, and the radial coordinate, respectively. The parameters $a$ and $b$ are rotation parameters
associated with the two independent angular momenta. We consider the equal-angular-momentum case $a = b$, which corresponds to the spinning Myers--Perry black hole \cite{Gibbons:2004ai,Gibbons:2004js}.

The calculations can be simplified by employing the following coordinate system \cite{Murata:2008xr}.
 \begin{equation}
\label{eqc111}
\begin{split}
t&=t_{H},\quad
 r^{2}=\frac{a^{2}+r_{H}^{2}}{1-\frac{a^{2}}{R^{2}}},\quad
\theta=2\theta_{H},\quad
 \phi=\phi_{H}-\psi_{H},\\
\psi&=-\frac{2at_{H}}{R^{2}}+\phi_{H}+\psi_{H},\quad
b=a,\quad
\mu=\frac{M}{(R^{2}-a^{2})^{3}}.
 \end{split}
\end{equation}

Eq.(\ref{eqc1}) can be rewritten in the following form
 \begin{equation}
\label{eqc2}
\begin{split}
ds^{2}=-\left(1+\frac{r^{2}}{R^{2}}\right)dt^{2}+\frac{dr^{2}}{G(\text{r)}}+\frac{r^{2}}{4}\left((\sigma^{1})^{2}+(\sigma^{2})^{2}+(\sigma^{3})^{2}\right)+\frac{2\mu}{r^{2}}\left(dt+\frac{a}{2}\sigma^{3}\right)^{2},
 \end{split}
\end{equation}
with
 \begin{equation}
\label{eqc3}
\begin{split}
&G(r)=1+\frac{r^{2}}{R^{2}}-\frac{2\mu(1-\frac{a^{2}}{R^{2}})}{r^{2}}+\frac{2\mu a^{2}}{r^{4}},\quad
\mu=\frac{r_{h}^{4}(R^{2}+r_{h}^{2})}{2R^{2}r_{h}^{2}-2a^{2}(R^{2}+r_{h}^{2})},\\
&\sigma^{1}=-\sin\psi d\theta+\cos\psi \sin\theta d\phi,\,
\sigma^{2}=\cos\psi d\theta+\sin\psi \sin\theta d\phi,\,
\sigma^{3}=d\psi+\cos\theta d\phi,
 \end{split}
\end{equation}
where
\begin{equation}
\label{eqc4}
\begin{split}
-\infty<t<\infty,\ r_{h}<r<\infty,\ 0\leq\theta<\pi,\ 0\leq\phi<2\pi,\ 0\leq\psi<4\pi.
 \end{split}
\end{equation}
To take the planar limit, we introduce the local coordinates
 \begin{equation}
\label{eqc5}
\begin{split}
&t=\tau,\quad
\frac{R}{2}(\phi-\pi)=x_{1},\quad
\frac{R}{2}\tan(\theta-\frac{\pi}{2})=x_{2},\\
&\frac{R}{2}(\psi-2\pi)=x_{3},\quad
r=\tilde{r},
 \end{split}
\end{equation}
where $(\tau, \widetilde{r}, x_1, x_2, x_3)$ represent the new coordinates, which can be rescaled by $\beta(\beta\rightarrow \infty)$.
\begin{equation}
\label{eqc6}
\begin{split}
\tau&\rightarrow\beta^{-1}\tau,\quad
x_{1}\rightarrow\beta^{-1}x_{1},\quad
x_{2}\rightarrow\beta^{-1}x_{2},\\
x_{3}&\rightarrow\beta^{-1}x_{3},\quad
\tilde{r}\rightarrow\beta\tilde{r},\quad
\tilde{r_{h}}\rightarrow\beta\tilde{r_{h}}.
 \end{split}
\end{equation}
This coordinate transformation yields the planar black brane solution \cite{Garbiso:2020puw}
\begin{equation}
\label{eqc7}
\begin{split}
ds^{2}=\frac{r^{2}}{R^{2}}\left(-d\tau^{2}+dx_{1}^{2}+dx_{2}^{2}+dx_{3}^{2}+\frac{r_{h}^{4}}{r^{4}\left(1-\frac{a^{2}}{R^{2}}\right)}\left(d\tau+\frac{a}{R}dx_{3}\right)^{2}\right)+\frac{R^{2}r^{2}}{r^{4}-r_{h}^{4}}dr^{2},
 \end{split}
\end{equation}
where $a$ is the boost parameter and $r_h$ denotes the horizon. Setting $a = 0$ recovers the Schwarzschild black brane.

To obtain a dimensionless form of Eq.(\ref{eqc7}), we introduce the dimensionless coordinates, $\tilde{\tau}=\frac{\tau}{R},\quad\tilde{x}_i=\frac{x_i}{R},\quad\tilde{z}=\frac{R}{r}$, the dimensionless rotation parameter $\tilde{a}=\frac{a}{R}$ and rescale the line element according to $d\tilde{s}^2=\frac{ds^2}{R^2}$. For notational simplicity, the tildes are omitted below,
\begin{equation}
\label{eqc8}
\begin{split}
ds^{2}=\frac{1}{z^{2}}\left(-d\tau^{2}+dx_{1}^{2}+dx_{2}^{2}+dx_{3}^{2}+\frac{z^{4}}{z_{h}^{4}(1-a^{2})}(d\tau+a dx_{3})^{2}\right)+\frac{z_{h}^{4}}{z^{2}(z_{h}^{4}-z^{4})}dz^{2}.
 \end{split}
\end{equation}

The metric in Eq. \eqref{eqc8} represents the local planar limit of the globally rotating Myers–Perry geometry. In this limit, the original angular motion appears as a local tangential plasma flow, and \(x_{3}\) denotes the local flow direction. The parameter \(a=\Omega R\)~\cite{Garbiso:2020puw} is dimensionless and characterizes the local boost or flow. The present approximation captures the local relative motion between the probe and the rotating plasma, whereas effects associated with the finite distance from the rotation axis, spatial velocity gradients, vorticity, centrifugal acceleration, and finite-size geometry are beyond the strict planar limit. All variables below are dimensionless quantities obtained by scaling with \(R\), with the rescaled symbols written without tildes.

The corresponding dimensionless Hawking temperature is
\begin{equation}
\label{eqb1}
T=\frac{\sqrt{1-a^{2}}}{z_{h}\pi },
\end{equation}

\section{Light quark energy loss in the spinning black hole background}\label{sec:03}
The purpose of this section is to investigate how the relative motion between an energetic light quark and the local plasma flow affects the quark stopping distance and instantaneous energy-loss rate. We consider propagation both transverse and parallel to the local flow direction and employ two complementary holographic descriptions: the falling-string approach for the stopping distance and the shooting-string approach for the instantaneous energy-loss rate. It should be noted that the calculation performed below is not a passive coordinate transformation of a single jet-medium configuration. For each value of $a$, we use the same boundary laboratory coordinates $(\tau,x_1,x_2,x_3)$ and, in the falling-string calculation, hold fixed the laboratory-frame jet energy, momentum magnitude, and propagation direction. In the ultrarelativistic limit, the transverse and parallel jet momenta are chosen as
\begin{equation}
  q_{\perp}^{\mu}=(\omega,|\vec{q}|,0,0),  \qquad  q_{\parallel}^{\mu}=(\omega,0,0,|\vec{q}|),  \qquad  \omega=|\vec{q}|.
  \label{eq:fixed_lab_jet_data}
\end{equation}

Throughout this work, the energetic probe is represented by a classical string configuration. The calculation describes the stopping and energy dissipation of an energetic quark excitation, but it does not introduce an explicit intrinsic-spin degree of freedom. Consequently, spin–rotation and spin–vorticity couplings are not included. The dependence on \(a\) should therefore be interpreted as the interaction of the probe with the local tangential flow, rather than as a calculation of spin-dependent dynamics in a vortical plasma.

\subsection{Falling string}

In this subsection, we apply the method developed in \cite{Arnold:2010ir,Arnold:2011qi} to study light-quark jet quenching in the spinning black-hole background. The key idea is to examine the stopping distance of an image jet induced by a massless source field. In the WKB approximation, this corresponds to a massless particle following a null geodesic. As the gauge field's wave packet falls into the horizon, the corresponding boundary jet dissipates and eventually loses its energy. Thus, the stopping distance is defined as the maximum distance the jet can travel before it stops.

In the WKB approximation, one assumes that the wave packet of the massless gauge field in the gravity dual is well localized in momentum space. This allows the wave function to be factorized as
\begin{equation}
A_j(\tau,z)=\exp\left[\frac{i}{\hbar}\left(q_kx_k+\int dzq_z\right)\right]\tilde{A}_j(\tau,z),
\end{equation}
where $q_k$ denotes the 4-momentum, which is conserved due to the translational invariance of the metric along the four boundary dimensions. $q_z$ is the momentum component in the holographic direction. $\tilde{A}_j(\tau,z)$ is a function that varies slowly with $\tau$ and $z$, and the indices $j,k$ label the four spacetime coordinates.

The metric in Eq.~\eqref{eqc8} describes the local tangential-flow sector of the globally rotating plasma, with \(x_3\) chosen along the local flow direction. This leads us to investigate the light-quark energy loss for two configurations. In the transverse case, the light quark propagates along $x_1$ or $x_2$. In the parallel case, it propagates along the local flow direction $x_3$.

Let us first consider the transverse case. Taking the classical limit $\hbar \to 0$, the wave packet follows a null geodesic.
\begin{equation}
0=ds^2=g_{ij}dx^idx^j+g_{zz}dz^2,
\end{equation}
yielding
\begin{equation}
\frac{dz}{d\zeta}=\frac{1}{\sqrt{g_{zz}}}\left[-g_{ij}\frac{dx^i}{d\zeta}\frac{dx^j}{d\zeta}\right]^{1/2},\label{n1}
\end{equation}
where $\zeta$ is an affine parameter along the trajectory. $g_{zz} = \frac{1}{z^2 D(z)}$ and $D(z) = 1 - \frac{z^4}{z_h^4}$. Since the metric is invariant under four-dimensional translations, the corresponding 4-momentum components are conserved. $g_{ij}\frac{dx^j}{d\zeta}$ is conserved and proportional to $q_i$, which leads to
\begin{equation}
\frac{dx^i}{d\zeta} \propto g^{ij}q_j.\label{n2}
\end{equation}
Dividing (\ref{n2}) by (\ref{n1}) then yields
\begin{equation}
\frac{dx^i}{dz}=\sqrt{g_{zz}}\frac{g^{ij}q_j}{\left(-q_kg^{kl}q_l\right)^{1/2}}.
\label{stop}
\end{equation}

To proceed, we compute the stopping distance. Substituting the metric (\ref{eqc8}) into (\ref{stop}) then yields the stopping distance in the transverse case in the spinning black hole background,
\begin{equation}
x_{\perp}=\int_0^{z_h} dz\frac{|\vec{q}|}
{\sqrt{(1+a^2F(z))\omega^2 - D(z)|\vec{q}|^2}},
\label{stop2}
\end{equation}
where $F(z) = \frac{z^4}{z_h^4(1-a^2)}$.  With $a = 0$, the above equation reduces to the SYM expression \cite{Arnold:2010ir,Arnold:2011qi}. Eq. (\ref{stop2}) indicates that the particle carries spatial momentum along $x_1$ or $x_2$.

In the parallel case, the particle carries spatial momentum along $x_3$. The corresponding stopping distance is
\begin{equation}
x_{\parallel}=\int_0^{z_h} dz \frac{\big(-aF(z)\omega + (1-F(z))|\vec{q}|\big)}
{D(z)\sqrt{\big[(1+a^2F(z))\omega^2 + 2aF(z)\omega|\vec{q}| - (1-F(z))|\vec{q}|^2\big]}}.
\label{stop1}
\end{equation}

\begin{figure}[t]
    \centering
    \includegraphics[width=0.45\textwidth]{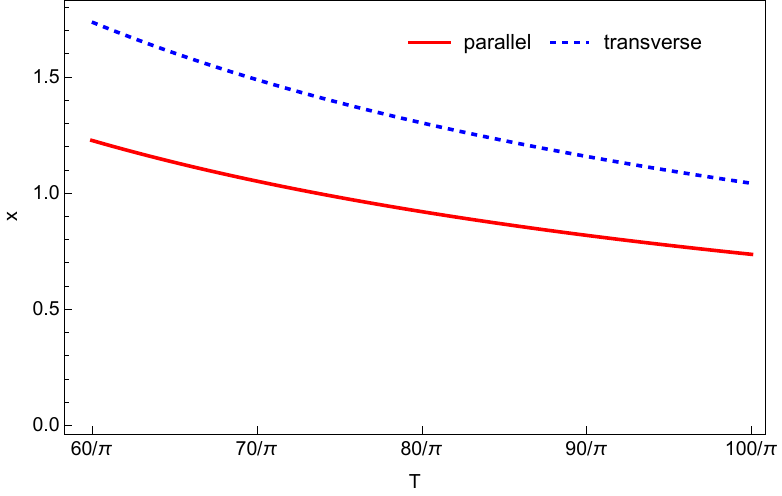}
    \includegraphics[width=0.45\textwidth]{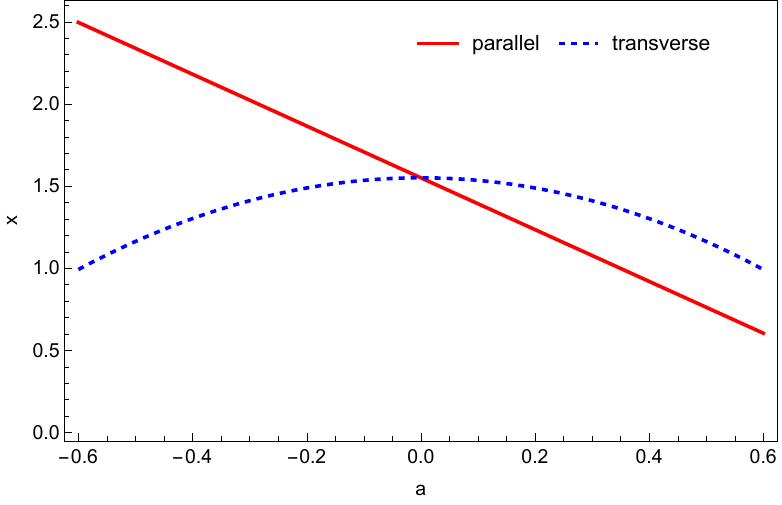}
    \caption{\textbf{Left panel:} Stopping distance $x$ as a function of temperature $T$ for $a=0.4$. \textbf{Right panel:} Stopping distance $x$ as a function of rotation parameter $a$ for $T=80/\pi$. }\label{fig1}
\end{figure}

As discussed in \cite{Garbiso:2020puw}, we take the planar black brane and a high temperature. Moreover, we take the rotation parameter to lie within the stable regime $|a| < 0.75$. We investigate the dependence of the light-quark stopping distance on the rotation parameter in the spinning black-hole background (dual to a rigidly rotating fluid). In numerical calculations, to keep the dimensionless initial radial position fixed during the scan over \(a\), we use the fixed lower and upper cutoff $z_0=\zeta_0z_h$, $z_{\rm{end}}=(1-\delta)z_h$ with $\zeta_0=8\times 10^{-3}, \delta=10^{-4}$.

The left panel of Fig.~\ref{fig1} displays the stopping distance of a light quark as a function of temperature $T$ for fixed rotation parameter $a$. It is observed that the stopping distance of light quarks decreases monotonically with increasing temperature, implying that they stop more readily in a hot environment. Meanwhile, the strength of the anisotropy induced by the rotation of the medium also weakens as the temperature rises.

The right panel of Fig.~\ref{fig1} displays the dependence on rotation parameter $a$ at fixed $T=80/\pi$. The transverse result is symmetric under $a\rightarrow-a$, reaches its maximum at $a=0$, and decreases as $\lvert a\rvert$ increases. By contrast, the parallel result is asymmetric because the parallel geodesic equation contains terms linear in $a$, originating from the off-diagonal metric component $g_{\tau x_3}$. For the ultrarelativistic probe with fixed \(z_0/z_h\) and \(z_{\rm end}/z_h\), the analytical derivation in appendix \ref{app} gives \(
x_{\parallel}=\frac{A-aB}{\pi T},\) where \(A\) and \(B\) are independent of \(a\). Thus, the parallel stopping distance depends linearly on \(a\), in agreement with the numerical results. In our setup, the probe propagates along the $+x_3$ direction. The parameter $a$ controls the local rotational flow. $a>0$ corresponds to a local medium flow along the negative $x_3$ direction and hence to a counter-propagating jet–medium configuration, leading to a reduced stopping distance. Negative $a$ corresponds to a co-propagating configuration and increases the parallel stopping distance. At $a=0$, the two curves coincide, as expected in the static isotropic limit. Therefore, in the co-propagating configuration, a light quark moving parallel to the medium flow travels farther before thermalizing than one moving transversely, whereas the opposite behavior occurs in the counter-propagating configuration.

\subsection{Shooting string}

In this subsection, we study the light-quark energy loss from the shooting string in the spinning Myers-Perry black hole background. In this approach \cite{Ficnar:2013wba,Ficnar:2013qxa}, a classical string endpoint starts near the horizon and travels toward the boundary, carrying finite energy and momentum that gradually bleed into the rest of the string.

Following \cite{Ficnar:2013wba,Ficnar:2013qxa}, we apply the shooting string method to the background metric (\ref{eqc8}). The resulting instantaneous energy loss of a light quark reads
\begin{equation}
\frac{dE}{dx^i}=\frac{1}{2\pi\alpha^\prime}\frac{g_{\tau\nu} \dot{X}^\nu}{\dot{x}^i},\label{de0}
\end{equation}
Here the dot denotes differentiation with respect to the parameter $\lambda$ along the endpoint worldline, $\dot X^\mu\equiv dX^\mu/d\lambda$, where $X^\mu=(\tau,x_1,x_2,x_3,z)$. In addition, we define the conserved ratio $L=\frac{E}{q_{x_i}}$ to characterize the null geodesic. In the subsequent numerical calculations, we set $1/(2\pi\alpha^\prime)=1$ for simplicity. A small $z$ (endpoint near the boundary) implies a large energy loss, which would quickly quench the jet and render it unobservable. Hence, we instead require the strings to start close to the horizon.

We consider two propagation channels defined by the conserved spatial momentum of the string endpoint. In the transverse channel, the nonzero momentum is directed along $x_1$, whereas the conserved momentum along the rotational-flow direction vanishes, $q_{3}=0$. In the parallel channel considered separately below, the endpoint carries nonzero momentum along $x_3$. For the transverse channel, the conserved energy and momenta of the endpoint are

\begin{equation}
E_{\perp}=-\frac{1}{\eta}\left(g_{\tau\tau} \dot{\tau}+g_{\tau x_3}\dot{x}_3\right),
\end{equation}
and
\begin{equation}
q_{x_1}=\frac{1}{\eta} g_{x_1x_1} \dot{x_1},\qquad q_{3}=\frac{1}{\eta} \left(g_{\tau x_3} \dot{\tau}+g_{x_3 x_3}\dot{x}_3\right)=0,
\end{equation}
where $\eta$ is the auxiliary field. $g_{\tau\tau}=-\frac{1-F(z)}{z^2}$, $g_{\tau x_3}=\frac{aF(z)}{z^2}$, $g_{x_3x_3}=\frac{1+a^2F(z)}{z^2}$ and $g_{x_1x_1}=\frac{1}{z^2}$. Thus, the null geodesics become
\begin{equation}
L_{\perp}=\frac{E_{\perp}}{p_{x_1}}=\frac{f(z)\dot{\tau}-aF(z)\dot{x}_3}{\dot{x}_1},
\end{equation}
where $f(z)=1-\frac{z^4}{z_h^4(1-a^2)}$.

Using the null condition $ds^2 = 0$, together with $q_{3}=0$, we obtain the transverse trajectory along $x_1$ and the accompanying induced drift along $x_3$:
\begin{equation}
\left(\frac{dx_1}{dz}\right)_{\perp}^2=\frac{1}{(1+a^2F(z))L_{\perp}^2-D(z)},\quad \left(\frac{dx_3}{dz}\right)_{\text{ind}}^2=\frac{a^2F(z)^2L_{\perp}^2}{D^2[(1+a^2F(z))L_{\perp}^2-D(z)]}. \label{dx}
\end{equation}

Because the metric contains the off-diagonal component $g_{\tau x_3}$, the condition $p_3=0$ does not imply $\dot{x}_3=0$. Instead, Eq.~(21) gives $\dot{x}_3=-\frac{g_{\tau x_3}}{g_{x_3x_3}}\,\dot{\tau}=-\frac{aF(z)}{1+a^2F(z)}\,\dot{\tau}$. Thus, an endpoint whose conserved spatial momentum is purely transverse may acquire an induced coordinate drift along the $x_3$ direction. This drift arises from the off-diagonal metric component and should not be confused with the independent parallel-propagation channel considered below.

At the radial turning point $z = z_*$, the common radial denominator in \eqref{dx} vanishes, yielding
\begin{equation}
L(z_*)_{\perp}=\sqrt{\frac{D(z_*)}{1+a^2F(z_*)}}. \label{nu0}
\end{equation}

Thus, the null trajectory in the transverse-momentum configuration becomes
\begin{equation}
\left(\frac{dx_1}{dz}\right)_{\perp}=-\sqrt{\frac{1+a^2F(z_*)}{F(z)-F(z_*)}},\quad \left(\frac{dx_3}{dz}\right)_{\text{ind}}=\frac{aF(z)}{D(z)}\sqrt{\frac{D(z_*)}{F(z)-F(z_*)}}.\label{nul}
\end{equation}

Only the first relation is required to evaluate the transverse energy-loss rate. The second relation records the induced longitudinal coordinate drift caused by the rotational background.

Combining (\ref{eqc8}), (\ref{de0}) and (\ref{nu0}) then yields the light quark energy loss
\begin{equation}
\left(\frac{dE}{dx_1}\right)_{\perp}=-\frac{L(z_*)_{\perp}}{2\pi\alpha^\prime z^2},\label{dex}
\end{equation}

We now turn to the independent parallel-propagation channel. In this case, the endpoint carries nonzero conserved momentum along $x_3$, while the transverse momenta vanish, $q_{1}=q_{2}=0$. One can obtain the null geodesics
\begin{equation}
L(z_*)_{\parallel} = \frac{-aF(z_*) + \sqrt{D(z_*)}}{1 + a^2F(z_*)},\label{numm}
\end{equation}
where $F(z_*) = \frac{z^4_*}{z_h^4(1-a^2)}$ and $D(z_*) = 1 - \frac{z_*^4}{z_h^4}$

The null geodesics equation in the parallel case is
\begin{equation}
\left(\frac{dx_3}{dz}\right)_{\parallel}=-\frac{1 - F(z) - a F(z) L(z_*)_{\parallel}}{D(z)\,\big(1 + a L(z_*)_{\parallel}\big)\,\sqrt{F(z) - F(z_*)}}.\label{nubb1}
\end{equation}

The light quark energy loss in the parallel case is
\begin{equation}
\left(\frac{dE}{dx_3}\right)_{\parallel}=-\frac{1}{2\pi\alpha^\prime z^2}\frac{L(z_*)_{\parallel}D(z)}{1-F(z)-aF(z)L(z_*)_{\parallel}},\label{dexb}
\end{equation}

To better understand the light-quark energy loss, we numerically integrate Eqs.~(\ref{nul}) and (\ref{nubb1}) in the small-$z_*$ limit and then invert the result to obtain $z(x)$. Then substitute into Eqs.~(\ref{dex}) and (\ref{dexb}), respectively. In asymptotically AdS$_5$ geometries, the shooting string approach usually takes $z_* \to 0$ \cite{Ficnar:2013wba,Ficnar:2013qxa}.

\begin{figure}[t]
    \centering
    \includegraphics[width=0.32\textwidth]{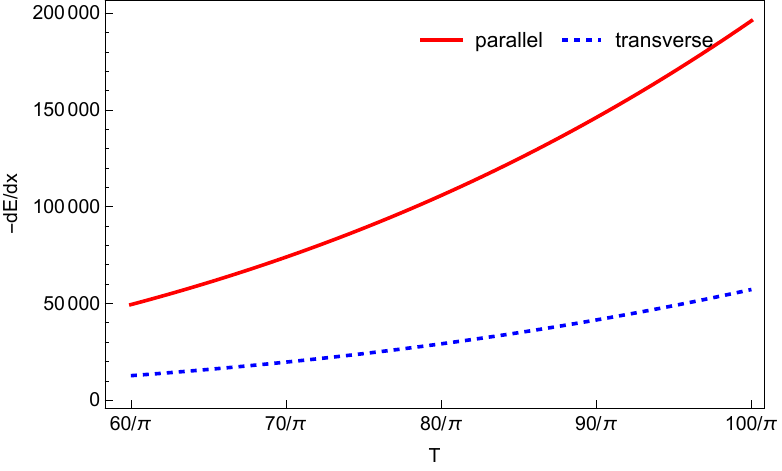}
    \includegraphics[width=0.32\textwidth]{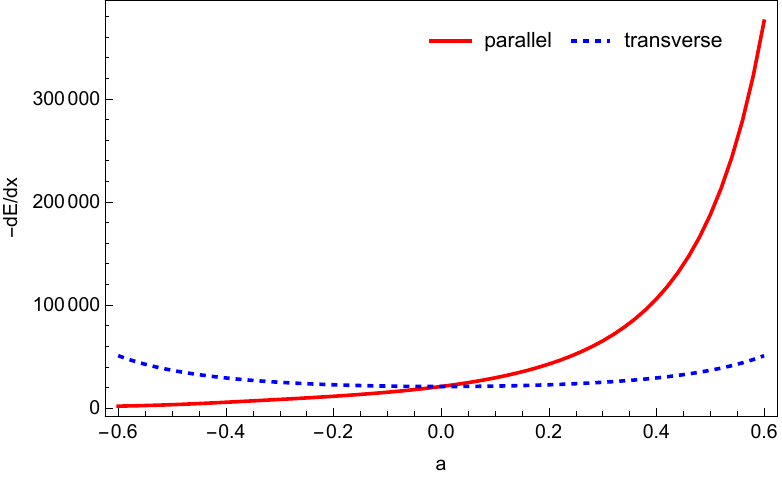}
    \includegraphics[width=0.32\textwidth]{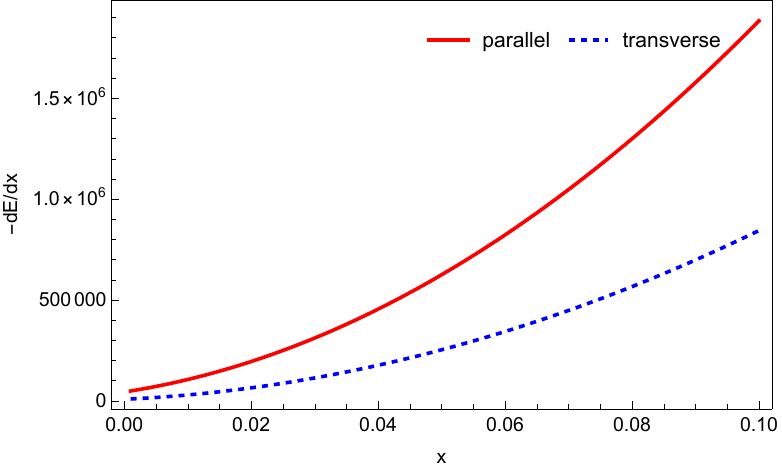}
    \caption{\textbf{Left panel:} Instantaneous energy loss $-dE/dx$ as a function of temperature $T$ for $a=0.4$ and $x=0.01$. \textbf{Middle panel:} Instantaneous energy loss $-dE/dx$ as a function of rotation parameter $a$ for $T=80/\pi$ and $x=0.01$. \textbf{Right panel:} Instantaneous energy loss $-dE/dx$ as a function of boundary spatial coordinate $x$ evaluated along the endpoint trajectory for $T=80/\pi$ and $a=0.4$.}\label{fig2}
\end{figure}

Fig.~\ref{fig2} summarizes the instantaneous energy loss of the shooting-string endpoint in the parallel and transverse propagations. The boundary coordinate is denoted by $x$, where $x=x_3$ for parallel propagation and $x=x_1$ for transverse propagation. In the left panel, $a=0.4$ and $x=0.01$ (The qualitative results are insensitive to the value of $x$) are held fixed. The energy-loss rates in both propagations increase monotonically with temperature, indicating that the endpoint loses energy more rapidly in a hotter medium.

The middle panel shows the dependence on the rotation parameter $a$ at fixed $T=80/\pi$ and $x=0.01$. The transverse result is symmetric under $a\rightarrow-a$, has its minimum at $a=0$, and increases with $\lvert a\rvert$. This symmetry follows from the fact that the transverse propagation depends on $a$ only through even combinations. In contrast, the parallel energy loss is strongly asymmetric and increases monotonically with $a$, owing to the terms linear in $a$ generated by the off-diagonal metric component $g_{\tau x_3}$. According to the convention $q_3>0$ adopted for the shooting-string endpoint, a counter-propagating jet-medium configuration ($a>0$) enhances the energy loss, whereas a co-propagating configuration ($a<0$) suppresses the parallel energy loss. The parallel rate grows more rapidly than the transverse rate because the parallel expression contains the additional $a$-dependent factor. Consequently, in the negative-$a$ region shown in the figure, the transverse energy-loss rate is larger than the parallel one, $(-dE/dx_1)_{\perp} > (-dE/dx_3)_{\parallel}$. For positive $a$, this ordering is reversed: the counter-propagating kinematics enhance the parallel energy-loss rate, leading to $(-dE/dx_3)_{\parallel} > (-dE/dx_1)_{\perp}$, with the difference between the two orientations becoming increasingly pronounced as $a$ increases.

The right panel presents the position dependence of the energy-loss rate at fixed $T=80/\pi$ and $a=0.4$. Here $x$ is the running boundary spatial coordinate evaluated along the shooting-string endpoint trajectory and should not be identified with the stopping distance $x$. As the endpoint propagates outward from the near-horizon region toward the boundary, its radial coordinate $z(x)$ decreases. Consequently, the factor $1/z^2(x)$ appearing in the energy-loss formulas produces a nonlinear increase of $-dE/dx$ with $x$.

These results are qualitatively consistent with those in Fig.~\ref{fig1} and show that both the local radial position of the string endpoint and the relative motion between the probe and the medium play essential roles in determining the instantaneous energy-loss rate.

\section{Conclusion and discussion}\label{sec:04}

In this work, we investigate light-quark energy loss in a rotating strongly coupled plasma by using holography. Both the falling-string and shooting-string descriptions have been employed. The gravitational background is obtained from the large-black-hole limit of the spinning Myers-Perry geometry and can be interpreted as a boosted/spinning black brane. In this setup, the parameter $a$ controls the strength of the boost/rotation effect of the dual
fluid.

Our results reveal a consistent physical picture from the two string descriptions. Increasing the temperature strengthens light-quark quenching: the falling-string analysis gives a shorter propagation distance before thermalization, while the shooting-string analysis yields a larger instantaneous energy-loss rate. For transverse motion, increasing the magnitude of the rotation parameter enhances energy loss. For parallel motion, however, the relative direction between the quark and the medium flow is decisive: a counter-propagating configuration enhances the energy loss and causes earlier stopping, whereas a co-propagating configuration suppresses the energy loss and allows the quark to propagate farther. In the static limit, the parallel and transverse results coincide, as required by isotropy. Moreover, the shooting-string energy loss increases nonlinearly as the endpoint moves from the near-horizon region toward the boundary. Thus, both approaches consistently show that stronger instantaneous energy loss corresponds to a shorter distance before thermalization.

From the boundary-fluid viewpoint, the observed directional dependence originates from the relative motion between the energetic quark and the local rotational flow. The present planar approximation only captures this local kinematic effect of rotation, while effects associated with finite-size geometry, velocity gradients, vorticity, and centrifugal acceleration~\cite{Chen:2023yug} require an analysis beyond the strict planar limit. Finally, the classical string description used here does not resolve the intrinsic spin of the light-quark excitation and does not include spin–rotation or spin–vorticity couplings. The results should therefore be understood as classical stopping and energy-dissipation results in the local planar limit. Our results are in qualitative agreement with previous holographic studies of rotation effects on heavy-quark dynamics \cite{Arefeva:2020jvo,Golubtsova:2021agl} and on the jet-quenching parameter in rotating plasmas \cite{Golubtsova:2022ldm,Zhu:2025bom}.

Several extensions are worth pursuing. In particular, it would be interesting to study light-quark energy loss in a rotating background with finite chemical potential, such as a charged rotating black hole background \cite{Cvetic:2004hs}. This would allow one to explore the combined effects of rotation and density on parton energy loss, which may be relevant for a more realistic description of strongly coupled QCD matter.

\section*{Acknowledgments}

Zhou-Run Zhu is supported by the Natural Science Foundation of Henan Province of China under Grant No. 242300420947, the startup Foundation projects for Doctors at Zhoukou Normal University, with the project number ZKNUC2023018. Yan-Qing Zhao is supported by the National Natural Science Foundation of China (NSFC) under Grant No. 12505151.

\appendix
\section{Analytical derivation of stopping distance in parallel case}\label{app}
\begingroup

For a probe propagating along the positive $x_3$ direction, we use the ultrarelativistic choice $q_3=\omega>0$. The radicand in the parallel trajectory equation (\ref{stop1}) then simplifies to
\begin{equation}
\mathcal{R}_{\parallel}
=(1+a^2F)\omega^2
  +2aF\omega q_3
  -(1-F)q_3^2
=\omega^2(1+a)^2F.
\end{equation}
The corresponding numerator becomes
\begin{equation}
-aF\omega+(1-F)q_3
=\omega\bigl[1-(1+a)F\bigr].
\end{equation}
Thus, for $|a|<1$, the cutoff-regulated parallel stopping distance is
\begin{equation}
x_{\parallel}
=
\int_{z_0}^{(1-\delta)z_h}
\frac{1-(1+a)F}
{D(1+a)\sqrt{F}}\,dz.
\end{equation}
Introducing $u=z/z_h$ and using
$z_h=\sqrt{1-a^2}/(\pi T)$, we obtain
\begin{equation}
\begin{aligned}
x_{\parallel}(T,a)
=\frac{1}{\pi T}
\int_{\zeta_0}^{1-\delta}
\frac{1-a-u^4}{u^2(1-u^4)}\,du
=\frac{1}{\pi T}
\left[
\int_{\zeta_0}^{1-\delta}\frac{du}{u^2}
-a\int_{\zeta_0}^{1-\delta}
\frac{du}{u^2(1-u^4)}
\right].
\end{aligned}
\end{equation}
Therefore,
\begin{equation}
x_{\parallel}(T,a)=\frac{A-aB}{\pi T}
\tag{R16}
\end{equation}
with
\begin{equation}
A=\frac{1}{\zeta_0}-\frac{1}{1-\delta},\qquad
B=\int_{\zeta_0}^{1-\delta}
\frac{du}{u^2(1-u^4)} =
\left[
-\frac{1}{u}
+\frac{1}{2}
\bigl(\operatorname{artanh}u-\arctan u\bigr)
\right]_{\zeta_0}^{1-\delta}.
\tag{R17}
\end{equation}
Since the integration limits are fixed and lie within $(0,1)$,
$A$ and $B$ are independent of $a$, and $B>0$. It follows that
\begin{equation}
\left.\frac{\partial x_{\parallel}}{\partial a}\right|_T
=-\frac{B}{\pi T}<0,
\qquad
\left.\frac{\partial^2x_{\parallel}}{\partial a^2}\right|_T=0.
\tag{R18}
\end{equation}
Hence, at fixed temperature, the parallel stopping distance is an
exact linear function of $a$ under the stated assumptions.

Note that the exact linear dependence established above
applies to the present ultrarelativistic calculation with fixed
$\zeta_0$ and $\delta$. It is not a general result for arbitrary
$\omega/|q_3|$ or other initial-position prescriptions.
\endgroup


\begin{thebibliography}{99}
\bibitem{Arsene:2004fa}
  I.~Arsene {\it et al.} [BRAHMS Collaboration],
  Nucl.\ Phys.\ A {\bf 757}, 1 (2005)
  doi:10.1016/j.nuclphysa.2005.02.130
  [nucl-ex/0410020].

\bibitem{Adcox:2004mh}
  K.~Adcox {\it et al.} [PHENIX Collaboration],
  Nucl.\ Phys.\ A {\bf 757}, 184 (2005)
  doi:10.1016/j.nuclphysa.2005.03.086
  [nucl-ex/0410003].

\bibitem{Back:2004je}
  B.~B.~Back {\it et al.},
  Nucl.\ Phys.\ A {\bf 757}, 28 (2005)
  doi:10.1016/j.nuclphysa.2005.03.084
  [nucl-ex/0410022].

\bibitem{Adams:2005dq}
  J.~Adams {\it et al.} [STAR Collaboration],
  Nucl.\ Phys.\ A {\bf 757}, 102 (2005)
  doi:10.1016/j.nuclphysa.2005.03.085
  [nucl-ex/0501009].

\bibitem{Wang:1992qdg}
X.~N.~Wang and M.~Gyulassy,
Phys. Rev. Lett. \textbf{68} (1992), 1480-1483
doi:10.1103/PhysRevLett.68.1480

\bibitem{Baier:1996kr}
R.~Baier, Y.~L.~Dokshitzer, A.~H.~Mueller, S.~Peigne and D.~Schiff,
Nucl. Phys. B \textbf{483} (1997), 291-320
doi:10.1016/S0550-3213(96)00553-6
[arXiv:hep-ph/9607355 [hep-ph]].

\bibitem{Majumder:2010qh}
A.~Majumder and M.~Van Leeuwen,
Prog. Part. Nucl. Phys. \textbf{66} (2011), 41-92
doi:10.1016/j.ppnp.2010.09.001
[arXiv:1002.2206 [hep-ph]].

\bibitem{Qin:2015srf}
G.~Y.~Qin and X.~N.~Wang,
Int. J. Mod. Phys. E \textbf{24} (2015) no.11, 1530014
doi:10.1142/S0218301315300143
[arXiv:1511.00790 [hep-ph]].

\bibitem{Shuryak:2004cy}
E.~V.~Shuryak,
Nucl. Phys. A \textbf{750} (2005), 64-83
doi:10.1016/j.nuclphysa.2004.10.022
[arXiv:hep-ph/0405066 [hep-ph]].

\bibitem{Kovtun:2004de}
P.~Kovtun, D.~T.~Son and A.~O.~Starinets,
Phys. Rev. Lett. \textbf{94} (2005), 111601
doi:10.1103/PhysRevLett.94.111601
[arXiv:hep-th/0405231 [hep-th]].









\bibitem{Maldacena:1997re}
J.~M.~Maldacena,
Adv. Theor. Math. Phys. \textbf{2}, 231-252 (1998)
doi:10.1023/A:1026654312961
[arXiv:hep-th/9711200 [hep-th]].

\bibitem{Witten:1998qj}
E.~Witten,
Adv. Theor. Math. Phys. \textbf{2}, 253-291 (1998)
doi:10.4310/ATMP.1998.v2.n2.a2
[arXiv:hep-th/9802150 [hep-th]].

\bibitem{Gubser:1998bc}
S.~S.~Gubser, I.~R.~Klebanov and A.~M.~Polyakov,
Phys. Lett. B \textbf{428}, 105-114 (1998)
doi:10.1016/S0370-2693(98)00377-3
[arXiv:hep-th/9802109 [hep-th]].

\bibitem{Casalderrey-Solana:2011dxg}
J.~Casalderrey-Solana, H.~Liu, D.~Mateos, K.~Rajagopal and U.~Achim Wiedemann,
Cambridge University Press, 2014,
ISBN 978-1-009-40350-4, 978-1-009-40349-8, 978-1-009-40352-8, 978-1-139-13674-7
doi:10.1017/9781009403504
[arXiv:1101.0618 [hep-th]].

\bibitem{DeWolfe:2013cua}
O.~DeWolfe, S.~S.~Gubser, C.~Rosen and D.~Teaney,
Prog. Part. Nucl. Phys. \textbf{75} (2014), 86-132
doi:10.1016/j.ppnp.2013.11.001
[arXiv:1304.7794 [hep-th]].

\bibitem{Liu:2006ug}
H.~Liu, K.~Rajagopal and U.~A.~Wiedemann,
Phys. Rev. Lett. \textbf{97} (2006), 182301
doi:10.1103/PhysRevLett.97.182301
[arXiv:hep-ph/0605178 [hep-ph]].

\bibitem{Liu:2006he}
H.~Liu, K.~Rajagopal and U.~A.~Wiedemann,
JHEP \textbf{03} (2007), 066
doi:10.1088/1126-6708/2007/03/066
[arXiv:hep-ph/0612168 [hep-ph]].

\bibitem{Chesler:2008uy}
P.~M.~Chesler, K.~Jensen, A.~Karch and L.~G.~Yaffe,
Phys. Rev. D \textbf{79} (2009), 125015
doi:10.1103/PhysRevD.79.125015
[arXiv:0810.1985 [hep-th]].

\bibitem{Arnold:2010ir}
P.~Arnold and D.~Vaman,
JHEP \textbf{10} (2010), 099
doi:10.1007/JHEP10(2010)099
[arXiv:1008.4023 [hep-th]].

\bibitem{Arnold:2011qi}
P.~Arnold and D.~Vaman,
JHEP \textbf{04} (2011), 027
doi:10.1007/JHEP04(2011)027
[arXiv:1101.2689 [hep-th]].

\bibitem{Chesler:2008wd}
P.~M.~Chesler, K.~Jensen and A.~Karch,
Phys. Rev. D \textbf{79} (2009), 025021
doi:10.1103/PhysRevD.79.025021
[arXiv:0804.3110 [hep-th]].

\bibitem{Ficnar:2013wba}
A.~Ficnar and S.~S.~Gubser,
Phys. Rev. D \textbf{89} (2014) no.2, 026002
doi:10.1103/PhysRevD.89.026002
[arXiv:1306.6648 [hep-th]].

\bibitem{Ficnar:2013qxa}
A.~Ficnar, S.~S.~Gubser and M.~Gyulassy,
Phys. Lett. B \textbf{738} (2014), 464-471
doi:10.1016/j.physletb.2014.10.016
[arXiv:1311.6160 [hep-ph]].

\bibitem{Zhu:2019ujc}
Z.~R.~Zhu, S.~Q.~Feng, Y.~F.~Shi and Y.~Zhong,
Phys. Rev. D \textbf{99} (2019) no.12, 126001
doi:10.1103/PhysRevD.99.126001
[arXiv:1901.09304 [hep-ph]].

\bibitem{Zhang:2023kzf}
Z.~q.~Zhang, X.~Zhu and D.~f.~Hou,
Eur. Phys. J. C \textbf{83} (2023) no.5, 389
doi:10.1140/epjc/s10052-023-11428-8

\bibitem{Zhang:2019gki}
Z.~q.~Zhang,
Eur. Phys. J. C \textbf{79} (2019) no.12, 992
doi:10.1140/epjc/s10052-019-7503-z

\bibitem{Zhang:2019jfq}
Z.~q.~Zhang,
Phys. Lett. B \textbf{793} (2019), 308-312
doi:10.1016/j.physletb.2019.04.064

\bibitem{Zhang:2025wxi}
L.~Zhang, L.~Yin, G.~D.~Zhou, C.~J.~Fan and X.~Chen,
Phys. Rev. D \textbf{111} (2025) no.12, 126001
doi:10.1103/m6rx-vy32
[arXiv:2504.04979 [hep-ph]].


\bibitem{Liang:2004ph}
Z.~T.~Liang and X.~N.~Wang,
Phys. Rev. Lett. \textbf{94} (2005), 102301
[erratum: Phys. Rev. Lett. \textbf{96} (2006), 039901]
doi:10.1103/PhysRevLett.94.102301
[arXiv:nucl-th/0410079 [nucl-th]].

\bibitem{Becattini:2007sr}
F.~Becattini, F.~Piccinini and J.~Rizzo,
Phys. Rev. C \textbf{77} (2008), 024906
doi:10.1103/PhysRevC.77.024906
[arXiv:0711.1253 [nucl-th]].


\bibitem{Baznat:2013zx}
M.~Baznat, K.~Gudima, A.~Sorin and O.~Teryaev,
Phys. Rev. C \textbf{88} (2013) no.6, 061901
doi:10.1103/PhysRevC.88.061901
[arXiv:1301.7003 [nucl-th]].

\bibitem{STAR:2017ckg}
L.~Adamczyk \textit{et al.} [STAR],
Nature \textbf{548} (2017), 62-65
doi:10.1038/nature23004
[arXiv:1701.06657 [nucl-ex]].

\bibitem{Jiang:2016woz}
Y.~Jiang, Z.~W.~Lin and J.~Liao,
Phys. Rev. C \textbf{94} (2016) no.4, 044910
[erratum: Phys. Rev. C \textbf{95} (2017) no.4, 049904]
doi:10.1103/PhysRevC.94.044910
[arXiv:1602.06580 [hep-ph]].

\bibitem{Chen:2020ath}
X.~Chen, L.~Zhang, D.~Li, D.~Hou and M.~Huang,
JHEP \textbf{07} (2021), 132
doi:10.1007/JHEP07(2021)132
[arXiv:2010.14478 [hep-ph]].

\bibitem{Zhao:2022uxc}
Y.~Q.~Zhao, S.~He, D.~Hou, L.~Li and Z.~Li,
JHEP \textbf{04} (2023), 115
doi:10.1007/JHEP04(2023)115
[arXiv:2212.14662 [hep-ph]].

\bibitem{Golubtsova:2021agl}
A.~A.~Golubtsova, E.~Gourgoulhon and M.~K.~Usova,
Nucl. Phys. B \textbf{979} (2022), 115786
doi:10.1016/j.nuclphysb.2022.115786
[arXiv:2107.11672 [hep-th]].


\bibitem{Arefeva:2020jvo}
I.~Y.~Aref'eva, A.~A.~Golubtsova and E.~Gourgoulhon,
JHEP \textbf{04}, 169 (2021)
doi:10.1007/JHEP04(2021)169
[arXiv:2004.12984 [hep-th]].

\bibitem{Golubtsova:2022ldm}
A.~A.~Golubtsova and N.~S.~Tsegelnik,
Phys. Rev. D \textbf{107} (2023) no.10, 106017
doi:10.1103/PhysRevD.107.106017
[arXiv:2211.11722 [hep-th]].



\bibitem{Hawking:1998kw}
S.~W.~Hawking, C.~J.~Hunter and M.~Taylor,
Phys. Rev. D \textbf{59}, 064005 (1999)
doi:10.1103/PhysRevD.59.064005
[arXiv:hep-th/9811056 [hep-th]].

\bibitem{Gibbons:2004ai}
G.~W.~Gibbons, M.~J.~Perry and C.~N.~Pope,
Class. Quant. Grav. \textbf{22}, 1503-1526 (2005)
doi:10.1088/0264-9381/22/9/002
[arXiv:hep-th/0408217 [hep-th]].

\bibitem{Gibbons:2004js}
G.~W.~Gibbons, H.~Lu, D.~N.~Page and C.~N.~Pope,
Phys. Rev. Lett. \textbf{93}, 171102 (2004)
doi:10.1103/PhysRevLett.93.171102
[arXiv:hep-th/0409155 [hep-th]].

\bibitem{Garbiso:2020puw}
M.~Garbiso and M.~Kaminski,
JHEP \textbf{12}, 112 (2020)
doi:10.1007/JHEP12(2020)112
[arXiv:2007.04345 [hep-th]].

\bibitem{Amano:2023bhg}
M.~A.~G.~Amano, C.~Cartwright, M.~Kaminski and J.~Wu,
Prog. Part. Nucl. Phys. \textbf{139} (2024), 104135
doi:10.1016/j.ppnp.2024.104135
[arXiv:2308.11686 [hep-th]].


\bibitem{Zhu:2024dwx}
Z.~R.~Zhu, S.~Wang, X.~Chen, J.~X.~Chen and D.~Hou,
Phys. Rev. D \textbf{110}, no.12, 126008 (2024)
doi:10.1103/PhysRevD.110.126008
[arXiv:2407.03633 [hep-ph]].

\bibitem{Zhu:2024uwu}
Z.~R.~Zhu, M.~Sun, R.~Zhou, Z.~Ma and J.~Han,
Eur. Phys. J. C \textbf{84}, no.12, 1252 (2024)
doi:10.1140/epjc/s10052-024-13628-2
[arXiv:2406.19661 [hep-ph]].

\bibitem{Zhu:2025ucq}
Z.~R.~Zhu, S.~Wang, Y.~K.~Liu and D.~Hou,
Phys. Rev. D \textbf{112}, no.2, 026012 (2025)
doi:10.1103/rwrh-z742
[arXiv:2501.04318 [hep-ph]].

\bibitem{Zhu:2025bom}
Z.~R.~Zhu, S.~Wang, M.~L.~Tian and D.~Hou,
Phys. Rev. D \textbf{113} (2026) no.6, 066014
doi:10.1103/s8dq-486v
[arXiv:2512.13076 [hep-ph]].


\bibitem{Murata:2008xr}
K.~Murata,
Prog. Theor. Phys. \textbf{121}, 1099-1124 (2009)
doi:10.1143/PTP.121.1099
[arXiv:0812.0718 [hep-th]].



\bibitem{Chen:2023yug}
J.~X.~Chen, D.~F.~Hou and H.~C.~Ren,
JHEP \textbf{03}, 171 (2024)
doi:10.1007/JHEP03(2024)171
[arXiv:2308.08126 [hep-ph]].

\bibitem{Cvetic:2004hs}
M.~Cvetic, H.~Lu and C.~N.~Pope,
Phys. Lett. B \textbf{598}, 273-278 (2004)
doi:10.1016/j.physletb.2004.08.011
[arXiv:hep-th/0406196 [hep-th]].



\end{thebibliography}
\end{document}